\documentclass[prl,aps,floatfix,amsmath,superscriptaddress,twocolumn, nofootinbib]{revtex4}
\usepackage{amssymb,enumerate}
\usepackage{graphicx}
\usepackage{graphics}
\usepackage{amsmath}
\usepackage{amsthm,bbm}
\usepackage{color}
\usepackage{dsfont}
\usepackage{hyperref}

\newtheorem{innercustomtheorem}{Theorem}
\newenvironment{customtheorem}[1]
  {\renewcommand\theinnercustomtheorem{#1}\innercustomtheorem}
  {\endinnercustomtheorem}
\newtheorem{innercustomlemma}{Lemma}
\newenvironment{customlemma}[1]
  {\renewcommand\theinnercustomlemma{#1}\innercustomlemma}
   {\endinnercustomlemma}
\newtheorem{innercustomproposition}{Proposition}
\newenvironment{customproposition}[1]
  {\renewcommand\theinnercustomproposition{#1}\innercustomproposition}
   {\endinnercustomproposition}

\usepackage{lipsum}
\usepackage{lmodern}
\usepackage{tcolorbox}

\usepackage{amsfonts}
\usepackage{graphicx,graphics,epsfig,times,bm,bbm,amssymb,amsmath,amsfonts,mathrsfs}
\usepackage[normalem]{ulem}
\usepackage{setspace}
\usepackage{subfigure}
\usepackage{dsfont}
\usepackage{braket}
\usepackage{upgreek }
\usepackage{tikz}
\usepackage{natbib}
\usepackage{chngcntr}
\usepackage{ulem}

\newtheorem{theorem}{Theorem}

\newtheorem{lemma}[theorem]{Lemma}

\newtheorem{proposition}[theorem]{Proposition}

\newcommand{\bes} {\begin{subequations}}
\newcommand{\ees} {\end{subequations}}
\newcommand{\bea} {\begin{eqnarray}}
\newcommand{\eea} {\end{eqnarray}}
\newcommand{\beq} {\begin{equation}}
\newcommand{\eeq} {\end{equation}}

\def\>{\rangle}
\def\<{\langle}
\def\Tr{\textrm{Tr}}

\newcommand{\ignore}[1]{}

\newcommand{\rob}[1]{{\color{black} #1}}

\begin{document}

\title{A no-broadcasting theorem for quantum asymmetry and coherence and a trade-off relation for approximate broadcasting}

\author{Iman Marvian}
\affiliation{Departments of Physics \& Electrical and Computer Engineering, Duke University, Durham, North Carolina 27708, USA}
\author{Robert W. Spekkens}
\affiliation{Perimeter Institute for Theoretical Physics, 31 Caroline St. N, Waterloo, \\
Ontario, Canada N2L 2Y5}


\begin{abstract}

Symmetries of both closed and open-system dynamics imply many significant constraints. These generally have instantiations in both classical and quantum dynamics (Noether's theorem, for instance, applies to both sorts of dynamics). We here provide an example of such a constraint which has no counterpart for a classical system, that is, a uniquely quantum consequence of symmetric dynamics. Specifically, we demonstrate
 the impossibility of broadcasting asymmetry (symmetry-breaking) relative to a continuous symmetry group, for bounded-size quantum systems. The no-go theorem states that if 
 two initially uncorrelated systems interact by symmetric dynamics and 
 asymmetry is created at one subsystem, then the asymmetry of the other subsystem must be reduced.  We also find a quantitative relation describing the tradeoff between the subsystems.  
These results cannot be understood in terms of additivity of asymmetry, because, as we show here, any faithful measure of asymmetry violates both sub-additivity and super-additivity.  
Rather, they must be understood as a consequence of an (intrinsically quantum) information-disturbance principle.
Our result also implies that if a bounded-size quantum reference frame for the symmetry group, or equivalently, a bounded-size reservoir of coherence (e.g., a clock with coherence between energy eigenstates in quantum thermodynamics) is used to implement any operation that is not symmetric, then the quantum state of the frame/reservoir is necessarily disturbed in an irreversible fashion, i.e., degraded.  
\end{abstract}

\maketitle

\noindent\emph{Introduction--}  Finding the consequences of symmetries of a closed or open quantum dynamics is 
a problem that has  a wide range of applications in physics, with Noether's theorem being perhaps the most prominent example. It is notable that the consequences that physicists have focussed on, including the conservation of Noether charges and currents, generally hold in {\em both} quantum and classical contexts. 
A natural question, therefore,  is whether there are consequences of symmetric dynamics that are unique to quantum theory.

Eugene P. Wigner pioneered the study of the consequences of symmetry in quantum theory and made various fundamental contributions to the topic.  For instance, in 1952,  he showed \cite{Wigner52, Wigner52-2} that under the restriction of using only Hamiltonians which conserve an observable $L$ that is additive across subsystems (e.g., the total angular momentum in a given direction), an exact measurement of another observable $O$ becomes impossible unless  $O$ commutes with $L$.
This fundamental no-go result, known as the Wigner-Araki-Yanase (WAY) theorem \cite{araki1960measurement, yanase1961optimal},  can equivalently be phrased as a consequence of the restriction to Hamiltonians which are invariant  under a  continuous symmetry, namely the symmetry for which $L$ is the generator.

In recent years, inspired by the success of entanglement theory \cite{horodecki2009quantum},  the problem of finding the consequences of symmetric dynamics  has been  further studied  in the framework of quantum resource theories \cite{horodecki2013quantumness, coecke2014mathematical, brandao2015reversible, chitambar2018quantum}.  In the resource theory of asymmetry \cite{footnote1},
any state which breaks the symmetry under consideration, i.e., any state which has some
 \emph{asymmetry}, is treated as a resource (similar to entangled states in entanglement theory). A particular case of interest, which is relevant in the context of the WAY theorem for instance, is when the symmetry under consideration is the continuous set of translations generated by a fixed observable $H$,
 i.e., $\{e^{-i H t}: t\in \mathbb{R}\}$ (Note that $H$ need not be the Hamiltonian, nor $t$ the time parameter, although the notation is meant to bring to mind this example). In this case a state contains asymmetry iff it contains \emph{coherence} (off-diagonal terms) with respect to the   
eigenspaces of $H$. It follows that the resource theory of asymmetry provides a natural framework to study this sort of coherence, which is 
known as \emph{unspeakable} coherence \cite{marvian2016quantify, streltsov2017colloquium}, and which is the notion that is relevant for quantum metrology~\cite{giovannetti2006quantum} and quantum thermodynamics \cite{ streltsov2017colloquium, lostaglio2015description, lostaglio2015quantumPRX} (as argued in Ref.~\cite{marvian2016quantify}). 

The resource-theoretic approach to the study of symmetric dynamics and asymmetry properties of quantum states has shed new light on earlier work. For instance, it was found that  the \emph{skew information},  a function introduced by Wigner and Yanase \cite{wigner1963information} as a replacement for the von-Neumann entropy in the presence of symmetry, is, in fact, 
a  measure of asymmetry
~\cite{Marvian_thesis, marvian2014extending, girolami2014observable}.  Also, it was found in Ref.~\cite{marvian2012information} that the WAY theorem can be understood as a corollary of a deep result in quantum information theory, known as the 
 \emph{no-programming} theorem
  ~\cite{Nielsen:97a,duvsek2002quantum,d2005efficient}. 


Another no-go theorem about continuous symmetries was uncovered in Ref.~\cite{marvian2013theory}:
the \emph{no-catalysis} theorem.   
This result concerns state conversions using operations which are covariant (symmetric) with respect to a compact connected Lie group, and states that if  the pure state conversion
 $\psi\rightarrow \phi$ is not achievable, then the catalyzed version of this same conversion,  $\psi\otimes \eta\rightarrow \phi\otimes \eta$, is also not achievable for any choice of pure catalyst state, $\eta$, in a finite-dimensional Hilbert space  \cite{No-Catal} (See also \cite{vaccaro2018coherence} for related observations). The question of whether there is nontrivial catalysis in the resource theory of asymmetry was inspired by the fact that there is nontrivial catalysis in entanglement theory \cite{JonathanPlenio, SandersGour} 

Taking the perspective of resource theories has also made evident that existing results on symmetric dynamics, including the no-catalysis and WAY theorems, are not uniquely quantum.  This is because it has clarified that a key assumption in each of these no-go theorems is that the resource state being used is not perfectly asymmetric in the sense that the state and its translated versions (under the symmetry transformations)
  are not perfectly distinguishable. If one makes the analogous assumption classically---that the probability distribution over classical configurations constituting one's resource 
 is not perfectly asymmetric in the same sense---
  then one obtains similar no-go results.

In this Letter, we find an example of a consequence of symmetric dynamics that {\em is} uniquely quantum, namely, a 
\emph{no-broadcasting theorem} for asymmetry. 
It asserts that if, during a symmetric dynamics on an initially uncorrelated pair of systems,  asymmetry is created
at one subsystem, then the asymmetry of another subsystem should
reduce.  We also show that this result does not hold classically.

In fact, we prove a more general result, 
namely, that under symmetric dynamics, if one uses a bounded-size quantum system in an asymmetric state (a reference frame or coherence reservoir) as a resource to perform an asymmetric operation (i.e., a task which is impossible under symmetric dynamics), then one necessarily disturbs its state irreversibly---the frame/reservoir degrades.
While it has been previously noted that quantum reference frames can  degrade when used to implement certain asymmetric operations~\cite{bartlett2006degradation,bartlett2007degradation,poulin2007dynamics,boileau2008quantum},  this conclusion was established only for certain target operations and considered only for the case where the frame starts in a pure state (See, however, \cite{cirstoiu2017irreversibility} for more recent work).



We also find a tradeoff relation for approximate broadcasting, namely,
 a lower bound on the amount of disturbance caused by the broadcasting of asymmetry/coherence in the case of pure states (Eq.~\eqref{Eq-trade}).  
This investigation also leads us to take note of a very general 
 constraint on measures of asymmetry
  (theorem \ref{Thm2}). 



\noindent\emph{Covariance condition--} 
We begin with some formalism.
Consider an arbitrary physical process 
 with input systems $Q$ and $S$ and output systems $Q'$ and $S'$, and let  $\Lambda_{QS\rightarrow Q'S'}$  (or simply $\Lambda$) be the corresponding quantum operation (i.e., completely positive trace-preserving linear map) 
  from the density operators of $QS$ to the density operators of $Q'S'$. We are interested in the processes satisfying the  \emph{covariance} condition
\begin{align}\label{cov-def4}
\forall t\in\mathbb{R}:\  \Lambda\circ [\mathcal{U}_Q(t)\otimes \mathcal{U}_S(t)]= [\mathcal{U}_{Q'}(t)\otimes \mathcal{U}_{S'}(t)]\circ \Lambda\ .
\end{align}
 Here, for each system $X\in\{Q, S, Q', S'\}$, we have defined $\mathcal{U}_X(t)[\cdot]\equiv e^{-i H_X t} (\cdot) e^{i H_X t}$, where $H_X$ is a (Hermitian) observable defined on system $X$.  Note that for each system $X$, the map  $\mathbb{R}\ni t\rightarrow \mathcal{U}_X(t)$  can be interpreted as a representation of a group of translations. Eq.~\eqref{cov-def4} means
 that  the description of the process $\Lambda_{QS\rightarrow Q'S'}$ is independent of which reference frame for translations one uses.

 A particular case of interest is when the operator $H_X$ is the Hamiltonian describing the closed-system dynamics of $X$,
so that $\mathcal{U}_X(t)$ represents evolution for time $t\in \mathbb{R}$.
  In this case, the tensor product form of $\mathcal{U}_Q(t)\otimes \mathcal{U}_S(t)$ (and $\mathcal{U}_{Q'}(t)\otimes \mathcal{U}_{S'}(t)$) reflects the fact that systems $Q$ and $S$ (and systems $Q'$ and $S'$)  are  not interacting with one another before (and after) the process $\Lambda$, and, therefore, can be treated as separate non-interacting subsystems. Then, the covariance condition in Eq.~\eqref{cov-def4}  means that the effect of the process $\Lambda$ on the inputs  $Q$ and $S$, does not depend on the time at which the process acts on these systems. This property is satisfied, for instance, by any thermal machine that interacts a system with thermal baths and with work reservoirs (batteries).

\noindent\emph{Asymmetry as a resource--}  A simple consequence of a process satisfying the covariance condition in Eq.~\eqref{cov-def4} is that it cannot generate {asymmetry}.  Suppose the input state $\rho_{QS}$ is \emph{symmetric} with respect to the  symmetry represented by $t\rightarrow \mathcal{U}_Q(t)\otimes \mathcal{U}_S(t)$, i.e.,
\beq
\forall t \in\mathbb{R}: \ \mathcal{U}_Q(t)\otimes\mathcal{U}_S(t) [\rho_{QS}]=\rho_{QS}\ .
\eeq
Note that this holds iff $\rho_{QS}$ is diagonal, or \emph{incoherent} relative to the eigenspaces  of $H_Q\otimes I_S+I_Q\otimes H_S$, where $I_S$ and $I_Q$ are the identity operators on $S$ and $Q$.  Then,  it can be easily seen that the covariance of process $\Lambda$ implies that incoherent states of the input systems are mapped to incoherent states of the output systems.  In this sense, asymmetry, or coherence, is a \emph{resource} which cannot be generated under covariant operations. Obviously, the physical interpretation of this resource depends on the nature of the symmetry.
  For instance, only for states that are asymmetric with respect to time-translations is a system useful as a clock and only for states that are asymmetric with respect to rotations is a system useful as a gyroscope.

 
Under the restriction to processes which satisfy the covariance condition in Eq.~\eqref{cov-def4}, having access to a resource of asymmetry allows one to perform operations that would otherwise be impossible. For any fixed state $\rho_Q$ of system $Q$, let $\mathcal{E}_{S\rightarrow S'}$ be the quantum operation
 from $S$ to $S'$ induced by the covariant operation $\Lambda_{QS\rightarrow Q'S'}$,   
\beq\label{Ded-reduced}
\mathcal{E}_{S \rightarrow S'}(\cdot)\equiv {\rm Tr}_{Q'} \left[ \Lambda_{QS \rightarrow Q'S'}(\rho_Q \otimes \cdot)\right]\ .
\eeq
(Note that  $Q$ and $S$ are assumed to be initially uncorrelated.) It can be easily seen that if $\rho_Q$ is a symmetric state, then the map $\mathcal{E}_{S \rightarrow S'}$ is covariant, i.e.,  
satisfies $\forall t\in\mathbb{R}:\ \mathcal{E}\circ \mathcal{U}_S(t) =  \mathcal{U}_{S'}(t)\circ \mathcal{E}$. On the other hand, using a state $\rho_Q$ which contains asymmetry,  we can implement a non-covariant operation
$\mathcal{E}_{S \rightarrow S'}$.  

For instance, if the process $\Lambda$ satisfies time-translation symmetry, then using an input system $Q$, whose state $\rho_Q$ contains asymmetry with respect to time translations (or equivalently, contains coherence relative to the energy eigenspaces),
 one can implement on $S$ operations which do not satisfy time-translation symmetry.
Therefore, for an agent who seeks to implement an operation at a particular time relative to some time standard (i.e., reference clock) but who lacks access to it, such a system can constitute a token of the standard, a quantum clock that is synchronized with the reference clock.


\noindent\emph{Irreversibility and Degradation}-- 
Suppose that there is a covariant process under which $\rho_Q \rightarrow \sigma_{Q'}$.
We say that the state conversion \ $\rho_Q\rightarrow \sigma_{Q'}$ is {\em reversible}  in the resource theory  if there exists a covariant process $\mathcal{R}_{Q'\rightarrow Q}$ which recovers  $\rho_Q$ from $\sigma_{Q'}$,  i.e., $\mathcal{R}_{Q'\rightarrow Q}(\sigma_{Q'})=\rho_Q$; otherwise, 
we say that  the state conversion 
 is irreversible and that the asymmetry of $\rho_Q$ is \emph{degraded} under the state conversion.

\noindent\emph{Asymmetry degradation theorem}--The following theorem shows that using a bounded-size system $Q$ in an asymmetric state $\rho_Q$ to implement a non-covariant operation  on system $S$ necessarily  degrades the asymmetry of $\rho_Q$.

\begin{theorem}[Asymmetry degradation]\label{main-theorem}
Let $Q$ be a system with a finite-dimensional Hilbert space, prepared in state $\rho_Q$. 
Suppose  system $S$,  initially uncorrelated with $Q$, interacts with system $Q$ via a covariant process $\Lambda_{QS \rightarrow Q'S'}$. Let $\mathcal{E}_{S \rightarrow S'}$, defined in Eq.~\eqref{Ded-reduced}, be the effective map which determines  how  the reduced state of output $S'$ depends on the state of $S$ (for a fixed $\rho_Q$). 
If $\mathcal{E}_{S \rightarrow S'}$ is not covariant,
 then, for some states of $S$ (including the completely mixed state) the 
 conversion from $\rho_{Q}$ to $\sigma_{Q'}$  is irreversible, i.e.,  state $\rho_Q$ cannot be recovered from state $\sigma_{Q'}$ via a covariant process  $\mathcal{R}_{Q'\rightarrow Q}$. (See Fig.~\ref{FigA}.) 
\end{theorem}
\begin{figure} [h]
\begin{center}
\includegraphics[scale=.5]{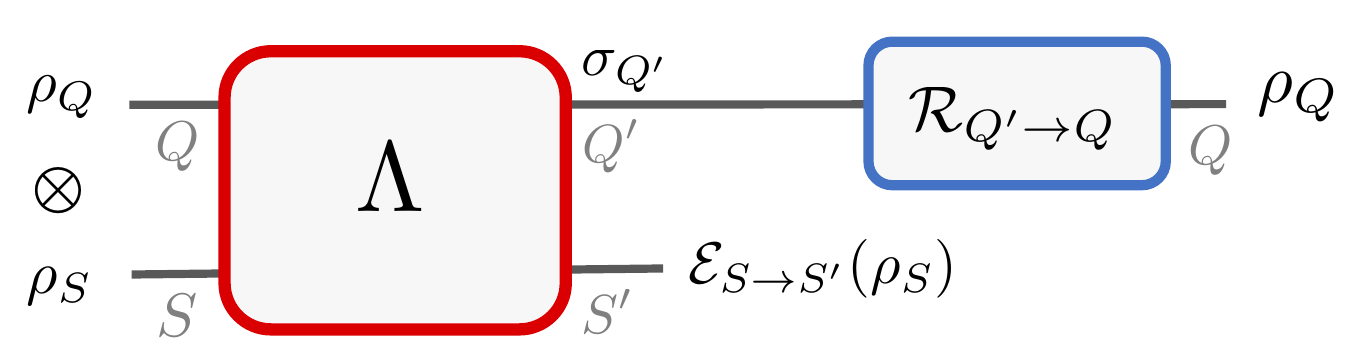}
\caption{ If, using a covariant operation $\mathcal{R}_{Q'\rightarrow Q}$, state $\rho_Q$ can be recovered from $\sigma_{Q'}$,  then the effective operation $\mathcal{E}_{S\rightarrow S'}$ is  covariant, and therefore can be implemented without having access to  
$\rho_Q$.
}
\label{FigA}
\end{center}
\end{figure}


The proof of this result will be given in the Supplementary Material (SM).  It leverages our result on no-broadcasting of asymmetry (proposition~\ref{prepos}), which is a special case of theorem~\ref{main-theorem}, and whose proof will be presented in this letter.

It is worth noting that, unlike the no-catalysis theorem of \cite{No-Catal}, here we do not assume that systems $Q$ and $S'$ are uncorrelated after the recovery process $\mathcal{R}_{Q'\rightarrow Q}$ is applied; rather, the result concerns 
  the reduced state of $Q$ itself. Such correlations become relevant, for instance, if we want to repeat this process to implement  $\mathcal{E}_{S \rightarrow S'}$ multiple times, i.e., to implement $\mathcal{E}^{\otimes n}_{S \rightarrow S'}$ for arbitrary integer $n$. As we see in the following, if one requires such a notion of repeatability, which amounts to assuming lack of correlations, then the proof of degradation
   becomes much simpler and can be achieved by using arguments similar to the no-catalysis theorem of \cite{No-Catal} or the arguments of \cite{vaccaro2018coherence}.  However, interestingly, according to  theorem \ref{main-theorem}, even if we relax this requirement and ignore correlations, 
   the degradation
    still holds, i.e.,  using  state $\rho_Q$ to implement a non-covariant process $\mathcal{E}_{S \rightarrow S'}$, will necessarily  imply that $Q$ undergoes a state conversion $\rho_{Q} \to \sigma_{Q'}$ that is irreversible.
  


\noindent\emph{No-broadcasting of asymmetry/coherence}-- 
The special case of Theorem \ref{main-theorem} that is the focus of this work concerns a map that incorporates both the process $\Lambda_{QS \rightarrow Q'S'}$ as well as any recovery operation $\mathcal{R}_{Q'\rightarrow Q}$ on it, and which is specialized to the case where $S$ is trivial.
We can conceptualize such a map  as a \emph{broadcast map} from $Q$ to the pair of systems $Q$ and $S'$. 
Unlike the usual discussions of broadcasting~\cite{barnum1996noncommuting}, where there is a set of possible states at the input and no restriction on the nature of the broadcast map, we are here interested in the case where there is a single state at the input, but the broadcast map is constrained to be covariant.

We will say that asymmetry/coherence can be \emph{broadcast} if there is a covariant map that takes any input state $\rho_Q$ to a state $\sigma_{QS'}$ with the property that 
\begin{enumerate}
\item[(i)] the input state $\rho_Q$ is reproduced in the output $Q$, i.e., $ \sigma_Q = \rho_Q$ where $\sigma_{Q} \equiv \Tr_{S'}(\sigma_{QS'})$, and 
\item[(ii)] the state of system $S'$ has nontrivial asymmetry/coherence, i.e., $\left[ \sigma_{S'}, H_{S'}\right]\ne 0$. 
\end{enumerate}

We can now state precisely the no-go result advertised in the title.


\begin{proposition}\label{prepos} 
(No-broadcasting of asymmetry/coherence)
 For bounded-size system $Q$, there does not exist a covariant broadcast map (defined by conditions (i) and (ii) above).
In other words,
\begin{align}\label{NBT}
\rho_Q\rightarrow \sigma_{QS'} \Longrightarrow\ &\text{NOT} \big( \sigma_{Q} =  \rho_Q \  \ \text{AND}  \ \ \left[ \sigma_{S'}, H_{S'}\right]\ne 0\big).
\end{align}
\end{proposition}

We prove this proposition later by appealing to a lemma that concerns the standard notion of broadcasting (lemma~\ref{lem-fixed}).

\color{black}

To see that this no-go result does not apply to classical asymmetry, it suffices to note that a map that clones any point distribution on a classical configuration space is covariant relative to any symmetry and consequently such a map achieves broadcasting of asymmetry when acted on any distribution that breaks the symmetry of interest.
 \color{black}




\noindent\emph{Non-additivity of  asymmetry}-- At first glance, it may appear that the impossibility of broadcasting asymmetry should follow from an intuitive idea, namely, that asymmetry might be a kind of extensive quantity, so that to create asymmetry in the  system $S'$ one needs to reduce the asymmetry of $Q$.
This intuition can be formalized using the notion of measures of asymmetry (See, e.g., \cite{gour2008resource, gour2009measuring, marvian2014extending, vac2008}): a function $f$  from states to real numbers is called a measure of asymmetry if (i)  it is non-increasing under covariant operations, i.e., $\rho_A\rightarrow \sigma_B$ implies $f(\rho_A)\ge f(\sigma_B)$,  and (ii) it vanishes on symmetric states. A measure of asymmetry is called  \emph{faithful} if it vanishes \emph{only} on symmetric states.  The Wigner-Yanase Skew information, $f(\rho_X)\equiv -\Tr([\sqrt{\rho_X}, H_X]^2)/2$, is an example of a faithful measure of asymmetry where $H_X$ is the generator of the symmetry (e.g., $H_X$ is the Hamiltonian if the symmetry is time translations).

A measure of asymmetry, $f$, is called   \emph{sub-additive} 
  if for any state $\sigma_{AB}$ of a composite system $AB$,  $f(\sigma_{AB})\le f(\sigma_{A})+f(\sigma_{B})$
     where $\sigma_A$ and $\sigma_B$ are the reduced states of $\sigma_{AB}$ on $A$ and $B$, respectively. 
     It is called \emph{super-additive} if $f(\sigma_{AB})\ge f(\sigma_{A})+f(\sigma_{B})$.
     A measure of asymmetry is called \emph{additive} if it is both  sub-additive and super-additive.
 
Suppose that there was even a single faithful super-additive measure of asymmetry, $f$.  In this case, $\rho_Q\rightarrow \sigma_{QS'} $ would imply that  $f(\rho_Q)\ge f(\sigma_{QS'})\ge  f(\sigma_{Q})+f(\sigma_{S'})$. Since $f$ is assumed to be faithful, if $\sigma_{S'}$ is not symmetric, then $f(\sigma_{S'})>0$, and we would be able to infer that $f(\rho_Q)> f(\sigma_{Q})$ and consequently that $\rho_Q\rightarrow \sigma_{Q}$ is irreversible, which would prove the impossibility of broadcasting asymmetry.


However, interestingly, as we show in the Supplemental Material, there is no such measure:
\begin{theorem}\label{Thm2}
A faithful measure of asymmetry is neither super-additive, nor sub-additive. 
\end{theorem}
It follows that the argument articulated above---wherein one seeks to justify no-broadcasting of asymmetry from super-additivity of asymmetry---is not sound. Indeed, the fact that our no-broadcasting result holds in spite of theorem \ref{Thm2} makes it more surprising. As we discuss in the SM, the failure of super-additivity can be shown using the fact that it is possible to create arbitrarily many systems in symmetry-breaking states starting from a single system in such a state, e.g., using an approximate cloner  (See also \cite{takagi2018operational} for a related result on skew information). 

It is worth noting that  some faithful measures of asymmetry, such as skew information, are additive {\em on product states}. 
Therefore, the argument articulated above {\em does} yield a proof of our no-broadcasting theorem, Eq.~\eqref{NBT}, for the special case where 
$\sigma_{QS'}= \sigma_{Q}\otimes \sigma_{S'}$.  However, to prove the theorem in the general case we need more powerful tools from quantum information theory.

\noindent\emph{Approximate broadcasting}-- 
Next, we derive a quantitive version of our no-broadcasting theorem.\color{black}  \ Specifically, we assume that there is a covariant process which transforms  $\rho_Q$ to  $\sigma_{QS'}$, 
and we seek to find a quantitative limit on the degree of success in broadcasting in terms of the amount of asymmetry (unspeakable coherence) in the initial state $\rho_Q$.  
We express our tradeoff relation in terms of
 (i) the degree of  irreversibility of the state conversion $\rho_Q\rightarrow \sigma_{Q}$  (where $\sigma_{Q}\equiv \Tr_{S'}(\sigma_{QS'})$) and (ii) the amount of asymmetry (unspeakable coherence) left in state $\sigma_{S'}$ (where $\sigma_{S'}\equiv \Tr_{Q}(\sigma_{QS'})$).
\color{black}


To  quantify the degree of irreversibility in a state conversion
 $\rho_Q\rightarrow \sigma_{Q}$,  we consider the minimum achievable infidelity in recovering the initial state $\rho_Q$ from the final state $\sigma_{Q}$, 
\beq
\text{irrev}(\rho_Q,\sigma_{Q})\equiv 1-\max_{\mathcal{R}}\ \text{Fid}^2\left(\rho_Q, \mathcal{R} (\sigma_{Q})\right),
 \eeq
where the maximization is over all covariant quantum operations.
Here, $\text{Fid}(\tau_1,\tau_2)\equiv \|\sqrt{\tau_1}\sqrt{\tau_2}\|_1$ is the  (Uhlmann) fidelity \cite{uhlmann1976transition, NielsenAndChuang, wilde2013quantum}. This definition implies that $ \text{irrev}(\rho_Q,\sigma_{Q})$ is between 0 and 1, and the state conversion
$\rho_Q\rightarrow \sigma_{Q}$ is reversible iff $\text{irrev}(\rho_Q,\sigma_{Q})=0$. 

To  quantify the asymmetry left in state $\sigma_{S'}$,
 we consider a measure of asymmetry defined in terms of the fidelity. For any $t\in \mathbb{R}$, define $f_t(\rho)\equiv1-\text{Fid}(\rho,e^{-i H t}\rho e^{i H t})=1- \|\sqrt{\rho}e^{-i H t} \sqrt{\rho } \|_1$. As we show in the SM, $f_t$  is a measure of asymmetry for any $t\in\mathbb{R}$, and it takes values in $[0,1]$. $f_t(\rho)$ quantifies how distinguishable $\rho$ is from $e^{-i H t}\rho e^{i H t}$. 
 

The trade-off relation we prove, unlike our no-broadcasting theorem, is limited to the case where the initial state 
 is pure, a fact which we denote by writing $\rho_Q=\psi_Q$.  Specifically, if $\psi_Q\rightarrow \sigma_{QS'}$, then
\beq\label{Eq-trade}
\forall  t\in\mathbb{R}:\ \ \    f_t(\sigma_{S'})\le 4 \frac{\sqrt{\text{irrev}(\psi_Q,\sigma_{Q})}}{1-f_t(\psi_Q)}\ .
\eeq

This tradeoff relation states that for any $t\in\mathbb{R}$, 
 the asymmetry of $\sigma_{S'}$, as quantified by $f_t$, 
 is upper bounded by a multiple of the degree of  irreversibility of the state conversion
 $\psi_Q\rightarrow \sigma_{Q}$, as quantified by $\sqrt{\text{irrev}(\psi_Q,\sigma_{Q})}$.
 Note that as $f_t(\psi_Q)$ increases,  the derived lower bound on $\text{irrev}(\psi_Q,\sigma_{Q})$  decreases (See also \cite{cirstoiu2017irreversibility} for related work).

The proof is given in the SM.  
There, we also demonstrate that  this tradeoff relation immediately implies our no-broadcasting theorem, Eq.~\eqref{NBT}, for the special case where the state $\rho_Q$ is pure. 

 Finally, we present the proof of our no-broadcasting theorem for asymmetry in the general case, where $\rho_Q$ may be mixed and 
 $\sigma_{QS'}$ may have correlations between $Q$ and $S'$. 

\noindent\emph{Proof of no-broadcasting of asymmetry/coherence}--
To prove proposition~\ref{prepos},  we make use of the following lemma
concerning broadcasting of an unknown state.   

\begin{lemma}(No-broadcasting of information encoded quantumly)
\label{lem-fixed} 
Let $\{\rho_Q^{(x)}\}_x$ be an arbitrary set of states of system $Q$. Suppose that under a quantum operation $\mathcal{E}_{Q\rightarrow QS'}$ (the putative broadcasting map), the state $\rho_Q^{(x)}$ is converted to the state $\sigma_{QS'}^{(x)}$ of systems $Q$ and $S'$. Assume that under this map, the state of system $Q$ is preserved at the output, so that the reduced state on output $Q$, defined as $\sigma_{Q}^{(x)} \equiv {\rm Tr}_{S'} (\sigma_{QS'}^{(x)})$ satisfies $\sigma_{Q}^{(x)}= \rho_Q^{(x)}$ for all states in the set $\{\rho_Q^{(x)}\}_x$.  In this case, the reduced state on the output $S'$, defined as $\sigma_{S'}^{(x)} \equiv {\rm Tr}_{Q} (\sigma_{QS'}^{(x)})$, can be obtained from the input state $\rho_{Q}^{(x)}$ by performing a projective measurement with projectors $\{\Pi_Q^{(\mu)}\}_\mu$ that commute with all states in the set $\{ \rho_Q^{(x)}\}_x$, followed by a state preparation which depends only on the outcome of this measurement.  That is, the reduced state of $S'$\color{black} is constrained to be of the form $\sigma_{S'}^{(x)}=\sum_\mu  p^{(x)}_{\mu} \sigma^{(\mu)}_{S'}$, where $p^{(x)}_{\mu}  = \Tr(\rho^{(x)}_Q \Pi_Q^{(\mu)})$ is the probability of obtaining the $\mu$ outcome, and where $\{\sigma^{(\mu)}_{S'}\}_\mu$ is an arbitrary set of states.
\end{lemma}


This is proven in the SM using a result of Koashi and Imoto~\cite{koashi2002operations}.  Note that one can conceptualize this result not only as a no-go for broadcasting  of quantumly encoded information, but also as a type of information-disturbance principle.  Specifically, it asserts that if the channel is disturbance-free on $Q$, then the only information about the identity of the unknown state of $Q$ that can be obtained from $S'$ is information that is encoded {\em classically} in $Q$.  In the case where the set of states is noncommuting, this implies a nontrivial and intrinsically quantum constraint on information gain. (See SM for further discussion.) 


Next, we leverage this lemma to prove proposition~\ref{prepos}.  We assume that the asymmetry at the input $Q$ is preserved in the output $Q$ and show that this implies that the state of $S'$ is symmetric. We assume, therefore, that there exists a covariant operation that achieves the conversion $\rho_Q \rightarrow \sigma_{QS'}$ such that $\sigma_{Q} = \rho_{Q}$.
We now note that, by virtue of its covariance, this operation 
also achieves the conversion  $\mathcal{U}_Q(t)[\rho_Q]\rightarrow \mathcal{U}_Q(t)\otimes \mathcal{U}_{S'}(t)[\sigma_{QS'}]$ for all $t\in \mathbb{R}$.
 Given the assumption that $\sigma_{Q} = \rho_{Q}$, the marginal on $Q$ of $\mathcal{U}_Q(t)\otimes \mathcal{U}_{S'}(t)[\sigma_{QS'}]$ is $\mathcal{U}_Q(t)[\rho_Q]$, and so an operation that converts an unknown state drawn from the set $\{ \mathcal{U}_Q(t)[\rho_Q]: t\in \mathbb{R}\}$ into the corresponding state in the set $\{ \mathcal{U}_Q(t)\otimes \mathcal{U}_{S'}(t)[\sigma_{QS'}]: t\in \mathbb{R}\}$ is precisely a broadcasting map satisfying the assumption of lemma~\ref{lem-fixed}, where $t$ plays the role of $x$.

We then infer from lemma~\ref{lem-fixed}  that under such a broadcasting map, the state of the output $S'$ must be prepared based on the outcome $\mu$ of a projective measurement $\{\Pi_Q^{(\mu)} \}_\mu$ on the input $Q$, where $\{ \Pi_Q^{(\mu)} \}_{\mu}$ is a complete set of orthogonal projectors that commute with all states in $\{ \mathcal{U}_Q(t)[\rho_Q]: t\in \mathbb{R}\}$.

\rob{The next step of the argument is where the restriction of scope to continuous symmetries occurs.}  
For continuous symmetries, we can consider the derivative with respect to the parameter $t$. Defining $\rho^{(t)}_Q\equiv \mathcal{U}_Q(t)[\rho_Q] =   e^{-i H_{Q} t}\rho_{Q} e^{i H_{Q} t}$, we have
\begin{equation}\nonumber
i\frac{d}{dt} \Tr(\rho_Q^{(t)} \Pi_Q^{(\mu)})=\Tr([\Pi_Q^{(\mu)},H_Q] \rho_Q^{(t)})=\Tr(H_Q [\rho_Q^{(t)}, \Pi_Q^{(\mu)}])\ ,
\end{equation}
where the first equality is 
 Ehrenfest's theorem, and the second equality follows from the cyclic property of the trace.  
 Recalling that $[\rho_Q^{(t)}\ , \Pi_Q^{(\mu)}]=0$ for all $t\in \mathbb{R}$ and for all $\mu$, it follows that $\Tr(\Pi_Q^{(\mu)} \rho_Q^{(t)})$ is independent of $t$. 
Because the probability distribution over $\mu$ induced by $\rho_Q^{(t)}$ is independent of $t$, the state of $S'$, which, as established above, can only depend on $\rho_Q^{(t)}$ via the mediary of $\mu$, is also independent of $t$.  This can be expressed as $e^{-i H_{S'} t}\sigma_{S'} e^{i H_{S'} t}= \sigma_{S'}$, or equivalently, as $[\sigma_{S'}, H_{S'}]=0$,  which concludes the proof. 

\noindent\emph{Conclusion}--- In this work, we have demonstrated a uniquely quantum
 constraint on the manipulation of asymmetry (equivalently, unspeakable coherence),  namely, that it cannot be broadcast.  Note that a similar result is found independently in Ref. \cite{lostaglio2019coherence}, which is published concurrently with this letter. 
We have also found a tradeoff relation which quantifies the amount of irreversibility in a covariant state conversion that achieves {\em approximate} broadcasting of 
 asymmetry/coherence. Furthermore, we showed that for bounded-size systems, asymmetry necessarily degrades with use. 
 
 It is worth noting that the constraints we have described here are generic to  symmetries described by connected Lie groups.  This is because any symmetry transformation in such groups is an element of  
 a one-parameter subgroup in the form of $e^{-i L x}$ for a generator $L$, and $x\in\mathbb{R}$, and covariance with respect to the original group implies covariance with respect to this subgroup.  
 
 The results are also {\em specific} to continuous symmetries in that they generally do {\em not} hold for discrete symmetries. This parallels the situation with the celebrated WAY theorem \cite{Wigner52, Wigner52-2, araki1960measurement, yanase1961optimal}, and the no-catalysis theorem of \cite{No-Catal}.

A broader question suggested by our work is: For which quantum resources theories is it impossible to broadcast a resourceful state using the free operations defined by that resource theory?  Ref.~\cite{piani2008no} can be seen as providing another example in addition to the one described here.
 

\color{black}



\noindent\emph{Acknowledgments}: We would like to thank an anonymous referee for a useful comment which simplified the proof of proposition \ref{prepos}.   This research was supported by Perimeter Institute for Theoretical Physics. Research at Perimeter Institute is supported by the Government of Canada through the Department of Innovation, Science and Economic Development Canada and by the Province of Ontario through the Ministry of Research, Innovation and Science.

\bibliography{Ref_2018}

\newpage

\onecolumngrid

\newpage

\maketitle
\vspace{-5in}
\begin{center}

\Large{Supplementary Material}
\end{center}



\tableofcontents

\onecolumngrid



\section{Proof of the constraint on broadcasting an unknown state  (Lemma 4 in the letter)}

We here show that lemma 4 from the main text is a simple corollary of the Koashi-Imoto theorem~\cite{koashi2002operations}.   We begin by repeating the lemma:

\begin{customlemma}{4} 
(No-broadcasting of information encoded quantumly)\label{lem-fixed0} 
Let $\{\rho_Q^{(x)}\}_x$ be an arbitrary set of states of system $Q$. Suppose that under a quantum operation $\mathcal{E}_{Q\rightarrow QS'}$ (the putative broadcasting map), the state $\rho_Q^{(x)}$ is converted to the state $\sigma_{QS'}^{(x)}$ of systems $Q$ and $S'$. Assume that under this map, the state of system $Q$ is preserved at the output, so that the reduced state on output $Q$, defined as $\sigma_{Q}^{(x)} \equiv {\rm Tr}_{S'} (\sigma_{QS'}^{(x)})$ satisfies $\sigma_{Q}^{(x)}= \rho_Q^{(x)}$ for all states in the set $\{\rho_Q^{(x)}\}_x$.  In this case, the reduced state on the output $S'$, defined as $\sigma_{S'}^{(x)} \equiv {\rm Tr}_{Q} (\sigma_{QS'}^{(x)})$, can be obtained from the input state $\rho_{Q}^{(x)}$ by performing a projective measurement with projectors $\{\Pi_Q^{(\mu)}\}_\mu$ that commute with all states in the set $\{ \rho_Q^{(x)}\}_x$, followed by a state preparation which depends only on the outcome of this measurement.  That is, the reduced state of $S'$ is constrained to be of the form $\sigma_{S'}^{(x)}=\sum_\mu  p^{(x)}_{\mu} \sigma^{(\mu)}_{S'}$, where $p^{(x)}_{\mu}  = \Tr(\rho^{(x)}_Q \Pi_Q^{(\mu)})$ is the probability of obtaining the $\mu$ outcome, and where $\{\sigma^{(\mu)}_{S'}\}_\mu$ is an arbitrary set of states.
\end{customlemma}


We use the form of the Koashi-Imoto theorem given in Hayden~\cite{hayden2004structure} 
(in fact, we use an amalgam of the latter's Theorem 10 and Proposition 11)
 with the notation modified to match the conventions of this article.  (This theorem is closely related to one obtained by Lindblad~\cite{lindblad1999general} in work generalizing the original no-broadcasting theorem of \cite{barnum1996noncommuting}. The work on information-preserving structures \cite{blume2010information} also derived a similar result.)  In the following, $\mathcal{H}_A$ denotes the Hilbert space of system $A$, and $\mathcal{B}(\mathcal{H}_A)$ denotes the space of bounded operators on $\mathcal{H}_A$.

\begin{customtheorem}{5}
\label{KoashiImoto}[Koashi-Imoto \cite{koashi2002operations}] 
Associated to the set of states $\{ \rho^{(x)}_Q\}_x$, there exists a decomposition of $\mathcal{H}_Q$ as
\beq\label{Sdec123}
\mathcal{H}_Q=\bigoplus_\mu \mathcal{H}_{Q_\mu^L}\otimes \mathcal{H}_{Q_\mu^R}\ ,
\eeq
into a direct sum of tensor products, such that
\begin{enumerate}
\item each of the states $\rho^{(x)}_Q$ decomposes as
\beq\label{Sdec12}
\rho^{(x)}_Q=\bigoplus_\mu p^{(x)}_\mu (\rho^{(x)}_{Q^L_\mu} \otimes \omega_{Q^R_\mu}) \ ,
\eeq 
where $\rho^{(x)}_{Q^L_\mu}$ is a state on $\mathcal{B}(\mathcal{H}_{Q_\mu^L})$, $\omega_{Q^R_\mu}$ is a state on $\mathcal{B}(\mathcal{H}_{Q_\mu^R})$ (which is independent of $x$) and $(p^{(x)}_\mu)_{\mu}$ is a probability distribution over $\mu$. 
\item \color{black} For every $\mathcal{E}_{Q}$ which leaves  all of the $\rho^{(x)}_Q$ invariant\color{black}
\beq
\forall \mu: \mathcal{E}_Q \circ \mathcal{P}_{Q_\mu} = {\rm id}_{Q^L_\mu} \otimes \mathcal{E}_{Q^R_\mu},
\eeq
 where $\mathcal{P}_{Q_\mu}(\cdot) :=\Pi_{\mu}(\cdot)\Pi_{\mu}$ is a projection operation on $\mathcal{B}(\mathcal{H}_Q)$ with $\Pi_{\mu}$ the projector onto $\mathcal{H}_{Q_\mu^L}\otimes \mathcal{H}_{Q_\mu^R}$, where ${\rm id}_{Q^L_\mu} $ is the identity operation on $\mathcal{B}(\mathcal{H}_{Q^L_\mu})$, and where $\mathcal{E}_{Q^R_\mu}$ is a quantum operation on $\mathcal{B}(\mathcal{H}_{Q^R_\mu})$ such that $\mathcal{E}_{Q^R_\mu}(\omega_{Q^R_\mu})=\omega_{Q^R_\mu}$.
\end{enumerate}
\end{customtheorem}

In addition to theorem~\ref{KoashiImoto}, our proof of lemma~\ref{lem-fixed0}  will use the following proposition, which concerns {\em universal} broadcasting, that is, broadcasting for the set of all possible quantum states of the input.  It asserts that if a universal broadcasting map is perfect for one output, it is trivial for the other.

\begin{customproposition}{6}\label{complementarity}
Consider any quantum operation $\mathcal{E}_{A \to SA}: \mathcal{B}(\mathcal{H}_A) \to  \mathcal{B}(\mathcal{H}_A\otimes \mathcal{H}_S)$, and let the marginal maps be denoted $\mathcal{E}_{A} \equiv {\rm Tr}_S\circ \mathcal{E}_{A \to SA}$ and $\mathcal{E}_{A \to S} \equiv {\rm Tr}_A\circ \mathcal{E}_{A \to SA}$.  
The following implication holds:
  if $\mathcal{E}_A$ is the identity channel, $\mathcal{E}_{A} = {\rm Id}_A$, then $\mathcal{E}_{A \to S}$ is an erasure channel, i.e., $\mathcal{E}_{A \to S} = \tau_S {\rm Tr}_A$ for some state $\tau_S$. 
\end{customproposition}

\begin{proof}
Proposition~\ref{complementarity} follows from the linearity of the map $\mathcal{E}_{A \to SA}$ by a proof that is similar to that of the no-cloning theorem.  Suppose that $\{ |\psi_k\rangle\langle \psi_k|_A \}_k$ is a complete basis of operators on $\mathcal{B}(\mathcal{H}_A)$ and consider their image under $\mathcal{E}_{A \to SA}$.  The assumption that $\mathcal{E}_{A} = {\rm Id}_A$ implies that the most general possibility is
\beq
\mathcal{E}_{A \to SA}(|\psi_k\rangle\langle \psi_k|_A) = |\psi_k\rangle\langle \psi_k|_A \otimes \tau^{(k)}_S,
\eeq
for some set of states $\{ \tau^{(k)}_S\}_k$.  Now consider any other state $|\psi\rangle_A \in \mathcal{H}_A$.  The projector $|\psi\rangle\langle \psi|_A$ admits of a decomposition
\beq
|\psi\rangle\langle \psi|_A = \sum_k \alpha_k |\psi_k\rangle\langle \psi_k|_A,
\eeq
for some real $\alpha_k$. 
Linearity of the map $\mathcal{E}_{A \to SA}$ implies
\bea\label{esa1}
\mathcal{E}_{A \to SA}(|\psi \rangle\langle \psi|_A) &=& \sum_k \alpha_k \mathcal{E}_{A \to SA}(|\psi_k\rangle\langle \psi_k|_A)\nonumber\\
&=& \sum_k \alpha_k  |\psi_k\rangle\langle \psi_k|_A\otimes \tau^{(k)}_S.
\eea
Meanwhile, the assumption that $\mathcal{E}_{A} = {\rm Id}_A$ implies that 
\bea\label{esa2}
\mathcal{E}_{A \to SA}(|\psi \rangle\langle \psi|_A) &=& |\psi \rangle\langle \psi|_A \otimes \tau_S,
\eea
for some $\tau_S$.
Given the linear independence of $\{|\psi_k\rangle \langle \psi_k|_A\}_k$, the only way that the expressions \eqref{esa1} and \eqref{esa2} can be consistent for an arbitrary choice of $|\psi\rangle_A$ is if the $\tau_S^{(k)}$ are independent of $k$.  This implies that $\mathcal{E}_{A \to SA}(\cdot) = {\rm Id}_A (\cdot) \otimes \tau_S$ and consequently that $\mathcal{E}_{A \to S}(\cdot) = \tau_S {\rm Tr}_A (\cdot)$.
\end{proof}

We are now in a position to prove lemma~\ref{lem-fixed0} .

\begin{proof}
 Item (1) of theorem~\ref{KoashiImoto} implies that, relative to the decomposition of $\mathcal{H}_Q$ associated to 
 the set $\{ \rho_Q^{(x)}\}_x$
  (this decomposition is described in Eq.~\eqref{Sdec123}), the state of the $Q^R_{\mu}$ subsystems is fixed, independent of the variable $x$, and therefore independent of the identity of the unknown state, while the state of the $Q^L_{\mu}$ subsystems do depend on the identity of the unknown state, as does the distribution over $\mu$.  

Now consider the broadcasting map $\mathcal{E}_{Q\to QS'}$, and recall that by assumption the marginal map on $Q$, denoted $\mathcal{E}_Q$, leaves each element of the set $\{ \rho_Q^{(x)}\}_x$ invariant. 

Define the map
\beq\label{dgg}
\mathcal{E}_{Q^L_{\mu} \to Q^L_{\mu} S' }(\cdot) \equiv {\rm Tr}_{Q^R_{\mu}} \circ \mathcal{E}_{Q\to QS'} (\cdot\otimes \omega_{Q^R_\mu})= {\rm Tr}_{Q^R_{\mu}} \circ\mathcal{P}_{Q_\mu}\circ  \mathcal{E}_{Q\to QS'}\circ\mathcal{P}_{Q_\mu} (\cdot\otimes \omega_{Q^R_\mu})
\eeq
where $\{\omega_{Q^R_\mu}\}_\mu$ are the  states defined in theorem~\ref{KoashiImoto}. Also, define the maps
\bea
\mathcal{E}_{Q^L_{\mu}  } &\equiv& {\rm Tr}_{S'} \circ \mathcal{E}_{Q^L_{\mu} \to Q^L_{\mu} S' }\nonumber\\
\mathcal{E}_{Q^L_{\mu} \to S'  } &\equiv& {\rm Tr}_{Q^L_{\mu}} \circ \mathcal{E}_{Q^L_{\mu} \to Q^L_{\mu} S' }. \eea
Item (2) of theorem~\ref{KoashiImoto} implies that 
\beq
\mathcal{E}_{Q^L_{\mu}} = {\rm Id}_{Q^L_{\mu}}.
\eeq
It then follows from lemma~\ref{complementarity} that
\beq\label{ryyy}
\mathcal{E}_{Q^L_{\mu} \to S'} = \sigma^{(\mu)}_{S'}\;{\rm Tr}_{Q^L_{\mu}} 
\eeq
for some state $\sigma^{(\mu)}_{S'}$.  

We conclude that
\begin{align}
\mathcal{E}_{Q\rightarrow S'}(\rho^{(x)}_Q)&\equiv  \Tr_{Q} \circ \mathcal{E}_{Q\rightarrow QS'}(\rho^{(x)}_Q)\\ &= \Tr_{Q} \circ \mathcal{E}_{Q\rightarrow QS'}\Big(\bigoplus_\mu p^{(x)}_\mu (\rho^{(x)}_{Q^L_\mu} \otimes \omega_{Q^R_\mu})\Big)\\ &=\sum_\mu p^{(x)}_\mu  \ \Big[\Tr_{Q} \circ \mathcal{E}_{Q\rightarrow QS'}(\rho^{(x)}_{Q^L_\mu} \otimes \omega_{Q^R_\mu})\Big]   \\ &=\sum_\mu p^{(x)}_\mu  \mathcal{E}_{Q^L_{\mu} \to S'}(\rho^{(x)}_{Q^L_\mu})\\&= \sum_\mu p^{(x)}_\mu\  \sigma^{(\mu)}_{S'}\ ,
\end{align}
where to get the second line we have used the decomposition in Eq.(\ref{Sdec12}), to get the fourth line we have used the definition of $\mathcal{E}_{Q^L_{\mu} \to S'}$ in Eq.(\ref{dgg}), and to get the last line we have used Eq.(\ref{ryyy}). Finally, note that 
\beq
p^{(x)}_\mu=\Tr(\Pi_\mu \rho^{(x)}_Q)\ .
\eeq
Therefore, we conclude that for any input stat $ \rho^{(x)}_Q$, the output state of $S'$, i.e. state $\mathcal{E}_{Q\rightarrow S'}(\rho^{(x)}_Q)$, can be obtained by first performing the projective measurement described by the projectors $\{\Pi_\mu\}_\mu$, and then preparing $S'$ in state $ \sigma^{(\mu)}_{S'}$, where $\mu$ is the the outcome of measurement. This concludes the proof. 

\color{black}

%
\end{proof}
  
 We can summarize the contents of lemma~\ref{lem-fixed0} 
  as follows.
The encoding of $x$ into the state of $Q$, $x \to \rho^{(x)}_{Q}$, defines an encoding of $x$ into $\mu$, namely, $x \to {\bf p}^{(x)}$, where 
${\bf p}^{(x)}$ is a distribution on $\mu$,
 and it defines an encoding of $x$ into each quantum subsystem $\mathcal{H}_{Q_\mu^L}$, namely,  $x\to \rho^{(x)}_{Q^L_{\mu}}$.
The encoding $x \to {\bf p}^{(x)}$ is a {\em classical} encoding of $x$.
  If the subsystem $\mathcal{H}_{Q_\mu^L}$ is nontrivial (i.e., if it is at least two-dimensional), then the encoding $x \to \rho^{(x)}_Q$ is a {\em quantum} encoding of $x$.  Lemma~\ref{lem-fixed0} states that
only the information about $x$ that is {\em classically encoded} in the input $Q$ can be broadcast to both the output $Q$ and the output $S'$, while the information about $x$ that is {\em quantumly encoded} in the input $Q$ cannot be broadcast to both $Q$ and $S'$.  Furthermore, if this quantumly encoded information is perfectly transmitted to the output $Q$, then it is absent from $S'$. {This is the sense in which it is appropriate to refer to lemma~\ref{lem-fixed0} as ``No-broadcasting of  information encoded  quantumly''.}

It follows that if the broadcast channel is such that the output $S'$ {\em does} contain information about $x$ that is quantumly encoded in the input $Q$,  then the state of $Q$  will be disturbed, i.e., it will not be preserved at the output.  It is in this sense that the result is aptly described as an  information-disturbance principle. 

Note that if the set of states $\{ \rho^{(x)}_{Q}\}_x$ defining the broadcasting task is a {\em commuting set}, then the states are jointly diagonalizable, and by choosing the projective measurement $\{ \Pi_{\mu}\}_{\mu}$ to be a measurement of the diagonalizing basis, {\em all} of the information about $x$ that is encoded in the input $Q$ is available to $\mu$ and thus also to $S'$.  In other words, for an encoding $x \to \rho^{(x)}_{Q}$ where  $\{ \rho^{(x)}_{Q}\}_x$ is a commuting set of states, {\em all} of the information about $x$ that is encoded in the input $Q$ is {\em classically encoded} (because a commuting set of states is equivalent to a set of probability distributions), and consequently all  of the information about $x$ that is encoded in the input $Q$ is available at {\em both} of the outputs of the broadcast channel.  Thus nontrivial constraints on broadcasting arise {\em only} for noncommuting sets of states and it is in this sense that 
such constraints on broadcasting are an intrinsically quantum phenomenon.


\section{Proof of the Asymmetry Degradation theorem (Theorem 1 in the letter)}

Proposition 2 in the paper, i.e., the no-broadcasting of asymmetry, is a special case of theorem 1, the asymmetry degradation theorem, 
where the system $S$ is trivial and the system $Q'$ is isomorphic to $Q$.  Here, we use this special case, which was proven in the main text, together with the result of Ref.~\cite{marvian2020coherence} to prove theorem 1
 in the general case.

We prove the contrapositive of the inference described in the statement of the theorem, namely, that if for all states of $S$, the transformation of $\rho_Q$ to $\sigma_{Q'}$ is reversible (in the sense that $\rho_Q$ can be recovered by a covariant process), then $\mathcal{E}_{S\to S'}$ is covariant.
Suppose that system $S$ is initially prepared in the maximally mixed state, $ I_S/d_S$, where $d_S$ is the dimension of the Hilbert space of $S$. Then, under channel $\Lambda_{QS\rightarrow Q'S'}$, the input state $\rho_Q\otimes I_S/d_S$ of the input systems $Q$ and $S$ is mapped to the state
\beq
\Lambda_{QS\rightarrow Q'S'}\left(\rho_Q\otimes \tfrac{1}{d_S}I_S\right),
\eeq
of systems $Q'$ and $S'$. 
By assumption, there exists a  covariant recovery channel $\mathcal{R}_{Q'\rightarrow Q}$, which recovers $\rho_Q$ from the reduced state of $Q'$, such that
\beq\label{sggl}
\mathcal{R}_{Q'\rightarrow Q}\circ \Tr_{S'}\circ \Lambda_{QS\rightarrow Q'S'}\left(\rho_Q\otimes \tfrac{1}{d_S}I_S\right)=\rho_Q\ ,
\eeq
where $\Tr_{S'}$ denotes 
the  partial trace on system $S'$.

Now suppose that system $S$ is initially prepared in a maximally entangled state with a purifying system $\overline{S}$, i.e., suppose that their joint state is  
\beq
|\Psi\rangle_{S\overline{S}}=\frac{1}{\sqrt{d_S}}\sum_{i=1}^{d_S} |ii\rangle_{S\overline{S}}\ ,
\eeq
 where $\{|i\rangle: i=1,\cdots, d_S \}$ is an orthonormal basis.  
Because the marginal state on $S$ of $|\Psi\rangle_{S\overline{S}}$ is $\tfrac{1}{d_S}I_S$, 
  Eq.~\eqref{sggl} implies that
\beq\label{sbbg}
\mathcal{R}_{Q'\rightarrow Q}\circ \Tr_{S' \overline{S}}\circ\Lambda_{QS\rightarrow Q'S'}(\rho_Q\otimes |\Psi\rangle\langle\Psi|_{S\overline{S}} ) =\rho_Q\ ,
\eeq
 where to simplify the notation we have suppressed the identity channel acting on the purifying system $\overline{S}$.

 Next, we consider the representation of the symmetry on system $\overline{S}$. We choose the representation such that the joint state $|\Psi\rangle_{S\overline{S}}$ is a symmetric state.  Let 
 \beq
 H_{\overline{S}}=-H^T_{{S}}\ ,
 \eeq
 where $T$ is transpose relative to the $\{|i\rangle: i=1,\cdots, d_S \}$ basis. Then, it can be easily seen that
 \beq
(e^{-i H_S t}\otimes e^{-i H_{\overline{S}} t}) |\Psi\rangle_{S\overline{S}}=|\Psi\rangle_{S\overline{S}}\ .
 \eeq
 In other words,  state $ |\Psi\rangle_{S\overline{S}}$ is symmetric with respect to 
 the  symmetry represented by unitaries $\{(e^{-i H_S t}\otimes e^{-i H_{\overline{S}} t}): t\in \mathbb{R} \}$.  
 
 Consider the channel $ \mathcal{T}_{Q\rightarrow QS'\overline{S}}$ from the input $Q$ to output $QS'\overline{S}$ defined by 
 \beq\label{Tdefn}
  \mathcal{T}_{Q\rightarrow QS'\overline{S}}(\cdot)\equiv \mathcal{R}_{Q'\rightarrow Q}\circ \Lambda_{QS\rightarrow Q'S'}((\cdot) \otimes |\Psi\rangle\langle\Psi|_{S\overline{S}} )\ ,
 \eeq
 where again we have suppressed the identity channels to simplify the presentation.

 Then, it can be easily seen that (i) the channel $\mathcal{T}_{Q\rightarrow QS'\overline{S}}$  is covariant, i.e., it satisfies 
 \beq
\forall t:\ \ \mathcal{T}_{Q\rightarrow QS'\overline{S}}\Big(e^{-i H_Q t}(\cdot) e^{i H_Q t}\Big)=[e^{-i H_{Q} t}\otimes e^{-i H_{S'} t} \otimes e^{-i H_{\overline{S}} t} ]\mathcal{T}_{Q\rightarrow QS'\overline{S}}(\cdot) [e^{i H_{Q} t}\otimes e^{i H_{S'} t} \otimes e^{i H_{\overline{S}} t} ]\ .
\eeq
and (ii) Eq.~\eqref{sbbg} implies that for the input state $\rho_Q$, $\mathcal{T}_{Q\rightarrow QS'\overline{S}}$ leaves the reduced state of $Q$ unchanged, 
\beq
 \Tr_{S'\overline{S}}\circ \mathcal{T}_{Q\rightarrow QS'\overline{S}}(\rho_Q)=\rho_Q\ .
 \eeq
Therefore, by applying proposition 2 of the letter (no-broadcasting of asymmetry),
 we conclude that the reduced state of $S'\overline{S}$, denoted by
  \begin{align}\label{taudefn}
\tau_{S'\overline{S}}\equiv \Tr_{Q}(\mathcal{T}_{Q\rightarrow QS'\overline{S}}(\rho_Q)) ,
\end{align}
is symmetric, i.e.,
\begin{align}\label{wmlp}
\Big[\tau_{S'\overline{S}}\ ,\ (H_{S'}\otimes I_{\overline{S}}+I_{S'}\otimes H_{\overline{S}}) \Big]= 0\ .
\end{align}
Recall that the channel $\mathcal{E}_{S \rightarrow S'}$ is defined as
 \beq\label{Edefn}
\mathcal{E}_{S \rightarrow S'}(\cdot)\equiv {\rm Tr}_{Q'} \left[ \Lambda_{QS \rightarrow Q'S'}(\rho_Q \otimes \cdot)\right]\ .
\eeq
By Eqs.~\eqref{taudefn} and ~\eqref{Tdefn}, we infer that 
\begin{align}
\tau_{S'\overline{S}}
&= \Tr_{Q}\circ  \mathcal{R}_{Q'\rightarrow Q}\circ \Lambda_{QS\rightarrow Q'S'}(\rho_Q \otimes |\Psi\rangle\langle\Psi|_{S\overline{S}} ),
\end{align}
and given Eq~\eqref{Edefn}, this implies that 
\begin{align}
\tau_{S'\overline{S}}
 &= \mathcal{E}_{S\rightarrow S'}\otimes \mathcal{I}_{\overline{S}}(|\Psi\rangle\langle\Psi|_{S\overline{S}})\ ,
\end{align}

where,  $\mathcal{I}_{\overline{S}}$ is the identity map on system $\overline{S}$. 

To complete the proof of the theorem, it remains only to show that the fact that the state $\tau_{S'\overline{S}}$ is symmetric  implies that the channel $\mathcal{E}_{S\rightarrow S'}$ is covariant.

This is implied by the following result from \cite{marvian2020coherence}: 


\begin{customlemma}{7}\label{lem177}
(From \cite{marvian2020coherence})
Consider the state
\beq
\tau_{S'\overline{S}}=  \mathcal{E}_{S\rightarrow S'}\otimes \mathcal{I}_{\overline{S}}(|\Psi\rangle\langle\Psi|_{S\overline{S}})\ ,
\eeq
where $\mathcal{I}_{\overline{S}}$ is the identity channel on system $\overline{S}$ and $|\Psi\rangle_{S\overline{S}}$ is the symmetric maximally entangled state on $S\overline{S}$ defined above.  Then, $\mathcal{E}_{S\rightarrow S'}$ is covariant, 
\beq
\forall t:\ \  \mathcal{E}_{S\rightarrow S'}(e^{-i H_S t}(\cdot) e^{i H_S t})=e^{-i H_{S'} t}\mathcal{E}_{S\rightarrow S'}(\cdot)e^{i H_{S'} t}\ ,
\eeq
if and only if the state  $\tau_{S'\overline{S}}$ is symmetric, 
\beq
\forall t:\ \ (e^{-i H_{S'} t}\otimes e^{-i  H_{\overline{S}} t}) \tau_{S'\overline{S}} (e^{i H_{S'} t}\otimes e^{i  H_{\overline{S}} t}) = \tau_{S'\overline{S}},
\eeq
a condition which is equivalent to $\tau_{S'\overline{S}}$ commuting with the Hamiltonian $H_{S'}\otimes I_{\overline{S}}+I_{S'}\otimes H_{\overline{S}}$.
\end{customlemma}

This completes the proof of the asymmetry degradation theorem (Theorem 1 in the letter).

For completeness, we present here the proof of lemma \ref{lem177}, which was originally proven in \cite{marvian2020coherence}. 

\subsection{Proof of lemma~\ref{lem177}}
First, assume that $\mathcal{E}_{S\rightarrow S'}$ is covariant. Then,
\begin{align}
(e^{-i H_S t}\otimes e^{-i H_{\overline{S}} t})\ \tau_{S'\overline{S}}\ (e^{i H_S t}\otimes e^{i H_{\overline{S}} t}) &=\ (e^{-i H_S t}\otimes e^{-i H_{\overline{S}} t}) \Big[\mathcal{E}_{S\rightarrow S'}\otimes \mathcal{I}_{\overline{S}}(|\Psi\rangle\langle\Psi|_{S\overline{S}})\Big]   (e^{i H_S t}\otimes e^{i H_{\overline{S}} t})  \\ &=\  \mathcal{E}_{S\rightarrow S'}\otimes \mathcal{I}_{\overline{S}}((e^{-i H_S t}\otimes e^{-i H_{\overline{S}} t}) |\Psi\rangle\langle\Psi|_{S\overline{S}} (e^{i H_S t}\otimes e^{i H_{\overline{S}} t}))  \\ &= \mathcal{E}_{S\rightarrow S'}\otimes \mathcal{I}_{\overline{S}}(|\Psi\rangle\langle\Psi|_{S\overline{S}})\\ &=\tau_{S'\overline{S}} \ ,
\end{align}
where to get the second line we have used the covariance of $\mathcal{E}_{S\rightarrow S'}$, and to get the third line we have used $(e^{-i H_S t}\otimes e^{-i H_{\overline{S}} t}) |\Psi\rangle_{S\overline{S}}=|\Psi\rangle_{S\overline{S}}\ $. Therefore, if $\mathcal{E}_{S\rightarrow S'}$ is covariant, then $ \tau_{S'\overline{S}}$ {is symmetric}.

Conversely, assume $\tau_{S'\overline{S}}$ {is symmetric},
which means 
\begin{align}\label{wehhhh}
[e^{-i H_{S'} t}\otimes e^{-i H_{\overline{S}} t}] \mathcal{E}_{S\rightarrow S'}\otimes \mathcal{I}_{\overline{S}}(|\Psi\rangle\langle\Psi|_{S\overline{S}})  [e^{i H_{S'} t}\otimes e^{i H_{\overline{S}} t}]  =\mathcal{E}_{S\rightarrow S'}\otimes \mathcal{I}_{\overline{S}}(|\Psi\rangle\langle\Psi|_{S\overline{S}})  \ . 
\end{align}
Recall that $(e^{-i H_S t}\otimes e^{-i H_{\overline{S}} t}) |\Psi\rangle_{S\overline{S}}=|\Psi\rangle_{S\overline{S}}\ $, which implies $(I_S\otimes e^{-i H_{\overline{S}} t}) |\Psi\rangle_{S\overline{S}}=(e^{i H_S t}\otimes I_{\overline{S}}) |\Psi\rangle_{S\overline{S}}\ $.
Using this fact, the left-hand side of Eq.~\eqref{wehhhh} can be rewritten as
\begin{align}
&[e^{-i H_{S'} t}\otimes e^{-i H_{\overline{S}} t}] [\mathcal{E}_{S\rightarrow S'}\otimes \mathcal{I}_{\overline{S}}(|\Psi\rangle\langle\Psi|_{S\overline{S}})]  [e^{i H_{S'} t}\otimes e^{i H_{\overline{S}} t}] \\
&= [e^{-i H_{S'} t}\otimes I_{\overline{S}}] \mathcal{E}_{S\rightarrow S'}\otimes \mathcal{I}_{\overline{S}}\Big( [I_S\otimes e^{-i H_{\overline{S}} t}]  |\Psi\rangle\langle\Psi|_{S\overline{S}}  [I_S\otimes e^{i H_{\overline{S}} t}]\Big)  [e^{i H_{S'} t}\otimes I_{\overline{S}}] \\ &= [e^{-i H_{S'} t}\otimes I_{\overline{S}}] \mathcal{E}_{S\rightarrow S'}\otimes \mathcal{I}_{\overline{S}}\Big( [e^{i H_S t} \otimes I_{\overline{S}}]  |\Psi\rangle\langle\Psi|_{S\overline{S}}   [e^{-i H_S t} \otimes I_{\overline{S}}]\Big)  [e^{i H_{S'} t}\otimes I_{\overline{S}}]\\ &= \mathcal{E}^{(t)}_{S\rightarrow S'}\otimes \mathcal{I}_{\overline{S}}\Big(|\Psi\rangle\langle\Psi|_{S\overline{S}} \Big)\ , 
\end{align}
where
\beq
\mathcal{E}^{(t)}_{S\rightarrow S'}(\cdot)\equiv e^{-i H_{S'} t}\Big[\mathcal{E}_{S\rightarrow S'}\Big(e^{i H_S t} (\cdot) e^{-i H_S t}\Big)\Big] e^{i H_{S'} t}\ .
\eeq

Therefore, Eq.~\eqref{wehhhh} can be rewritten as 
\begin{align}
\mathcal{E}^{(t)}_{S\rightarrow S'}\otimes \mathcal{I}_{\overline{S}}(|\Psi\rangle\langle\Psi|_{S\overline{S}}) =\mathcal{E}_{S\rightarrow S'}\otimes \mathcal{I}_{\overline{S}}(|\Psi\rangle\langle\Psi|_{S\overline{S}})  \ . 
\end{align}
Given that $|\Psi\rangle_{S\overline{S}}=\frac{1}{\sqrt{d_s}}\sum_{i=1}^{d_S}|ii\rangle_{S\overline{S}}$, is a maximally entangled state, this immediately implies 
\begin{align}
\forall t\in\mathbb{R}:\ \ \mathcal{E}^{(t)}_{S\rightarrow S'}=\mathcal{E}_{S\rightarrow S'} \ , 
\end{align}
which is equivalent to the covariance condition 
\beq
\forall t\in\mathbb{R}:\ \ \mathcal{E}_{S\rightarrow S'}(e^{-i H_S t}(\cdot) e^{i H_S t})=e^{-i H_{S'} t}\mathcal{E}_{S\rightarrow S'}(\cdot)e^{i H_{S'} t}\ .
\eeq
This completes the proof.

\color{black}

\section{Proof that a faithful measure of asymmetry is neither super-additive, nor sub-additive (theorem 3 in the {letter}) }


To prove that a faithful measure of asymmetry is not sub-additive, we note that for any symmetry, theres exists states $\sigma_{AB}$ of the composite system $AB$, such that $\sigma_{AB}$  is asymmetric, while the reduced states of $A$ and $B$ can both be symmetric. For instance, $\sigma_{AB}$ can be a maximally entangled state that is asymmetric, while $\sigma_A$ and $\sigma_B$ are completely mixed states, and therefore symmetric.  The assumption of faithfulness of the measure of asymmetry $f$ implies that $f(\sigma_{AB})>0$, while $f(\sigma_A)=f(\sigma_B)=0$. It follows that $f(\sigma_{AB})>f(\sigma_A)+f(\sigma_B) $, implying that $f$ is not sub-additive.

Note that  the presence of entanglement is not necessary to observe a violation of sub-additivity. For instance, for any finite group $G$, we can define the state 
\beq
\sigma_{AB}=\frac{1}{|G|}\sum_{g\in G} |g\rangle\langle g|_A\otimes U_B(g) \rho_B U_B^\dag(g)\ ,
\eeq
 where  $G\ni g\rightarrow U_B(g)$  is the representation of symmetry on system $B$, and $\{|g\rangle: g\in G\}$ is a set of orthonormal states on $A$. We assume the representation of symmetry on $A$ is trivial. Then, it can be easily seen that if state $\rho_B$ is asymmetric, then the state $\sigma_{AB}$ will also be asymmetric. However, the reduced state on $B$, which is given by $\sigma_B=\frac{1}{|G|}\sum_{g\in G} U_B(g) \rho_B U_B^\dag(g)$, is {symmetric}.
 The reduced state of $A$ {is also symmetric},
  because the representation of symmetry on $A$ is trivial. It follows that for any measure of asymmetry $f$, $f(\sigma_A)=f(\sigma_B)=0$. However, since $\sigma_{AB}$ is asymmetric and $f$ is faithful, $f(\sigma_{AB})>0$. This implies that $f(\sigma_{AB})> f(\sigma_A)+f(\sigma_B)$, 
 so we have a violation of sub-additivity despite the fact that $\sigma_{AB}$ is unentangled.
   This argument can be easily generalized to the case of continuous groups as well.

To prove that a faithful measure of asymmetry is not super-additive, we use the fact that starting with any asymmetric state, a covariant map  can \emph{distribute} its asymmetry to an arbitrary number $n$ of systems, such that the reduced state of each system is a fixed (i.e., independent of $n$) asymmetric state. Specifically, one can consider the  universal cloner map \cite{keyl1999optimal},  which approximately clones its input state in a $d$-dimensional Hilbert space $A$ to arbitrary many output systems $A_1\cdots A_n$ with Hilbert spaces identical with $A$ (The explicit form of this map is presented in the footnote  \footnote{Explicitly the universal cloner map of \cite{keyl1999optimal} is given by 
$$\mathcal{E}_{A\rightarrow A_1\cdots A_n}(\rho)=\frac{d}{d(n)}\  \Pi_\text{sym}(\rho\otimes I^{\otimes (n-1)})\Pi_\text{sym}\ ,$$ 
 where $d$ is the dimension of $\mathcal{H}_A$, the Hilbert space of $A$, $d(n)={{d+n-1}\choose{n}}$ is the dimension of the symmetric subspace $\mathcal{H}^{\otimes n}_A$, and $\Pi_\text{sym}$ is the projector to this subspace (This is a special case of the universal cloner map of  \cite{keyl1999optimal}, where at the input we only have a single copy of $\rho$). Interestingly, it turns out that this map can be understood as a Petz recovery map \cite{marvian2016clocks, lemm2017information}). Applying this map to input state $\rho$ of system $A$, we obtain a state of $A_1,\cdots, A_n$ which is permutationally  invariant, and therefore, the reduced state of all subsystems are identical. In particular, for any $i\in \{1,\cdots, n\}$, the reduced state of system $A_i$ is
$$\rho_{A_i}=\Tr_{\overline{A_i}}\Big(\mathcal{E}_{A\rightarrow A_1\cdots A_n}(\rho)\Big)=c_n \rho+ [1-c_n] \frac{I}{d} $$
where the partial trace is over all systems $A_1,\cdots, A_n$ except system $A_i$, and 
$$c_n=\frac{d+n}{n(d+1)}\ .$$
 Note that as $n$ goes to infinity, $c_n$ converges to $c_\infty=(d+1)^{-1}$, which is strictly larger than zero.
}).

The universal cloner $\mathcal{E}_{A\rightarrow A_1\cdots A_n}$ of \cite{keyl1999optimal} is covariant with respect to SU(d) symmetry, i.e., for any unitary $U$, it satisfies   
\beq
\mathcal{E}_{A\rightarrow A_1\cdots A_n}(U(\cdot)U^\dag)=U^{\otimes n}\mathcal{E}_{A\rightarrow A_1\cdots A_n}(\cdot) {U^\dag}^{\otimes n}\ .
\eeq
Therefore, it also satisfies the covariance condition for any symmetry group. 

Let $\sigma_{A_1\cdots A_n}=\mathcal{E}_{A\rightarrow A_1\cdots A_n}(\rho_A)$ be the joint state of $A_1\cdots A_n$ for input $\rho_A$ and  
\beq
\sigma_{A_i}=\Tr_{\overline{A_i}}(\sigma_{A_1\cdots A_n})
\eeq
 be the reduced state of  system $A_i$, where the partial trace is over all subsystems except $A_i$. Then, the covariance  of the universal cloner implies that for any input state $\rho_A$, the reduced state of each system $A_i$  is in the form 
\beq
\sigma_{A_i}=c_n \rho_A+ (1-c_n) \frac{I}{d}\ ,
\eeq
where $0<c_n<1$  determines the fidelity of cloning  and is bounded away from $0$, even in the limit $n\rightarrow \infty$ \cite{keyl1999optimal}.  Therefore, if state $\rho_A$ breaks a given symmetry, then state $\sigma_{A_i}$ will also break that symmetry. 

Suppose $f$ is a measure of asymmetry. Then, the covariance of the universal cloner  $\mathcal{E}_{A\rightarrow A_1\cdots A_n}$ implies
\beq
f(\sigma_{A_1\cdots A_n})=f(\mathcal{E}_{A\rightarrow A_1\cdots A_n}(\rho_A))\le f(\rho_A)\ .
\eeq
If $f$ is super-additive then
\beq
f(\sigma_{A_1\cdots A_n})\ge \sum_{i=1}^n f(\sigma_{A_i})= n\times f(c_n \rho_A+ (1-c_n) \frac{I}{d}) .
\eeq
Combining these two inequalities, we conclude that
\beq\label{ertyu}
f(\rho_A)\ge n\times f(c_n \rho_A+ (1-c_n) \frac{I}{d})  \ .
\eeq
In the limit that $n$ goes to infinity $c_n$ converges to a fixed non-zero constant $c_\infty$ (For the universal cloner map of \cite{keyl1999optimal}, $c_n=\frac{d+n}{n(d+1)}$, and therefore $c_\infty=(d+1)^{-1}>0$.  See the footnote for further details).  Therefore,  the state $c_\infty \rho_A+ (1-c_\infty) \frac{I}{d}$ still breaks the symmetry .  Therefore, if $f$ is faithful then $f(c_\infty \rho_A+ (1-c_\infty) \frac{I}{d})$ is strictly larger than zero. But this implies that, in the limit that $n$ goes to infinity, the right-hand side of  Eq.~(\ref{ertyu}) diverges, while the left-hand side remains finite. This leads to a contradiction and proves that a faithful measure of asymmetry cannot be  super-additive.  

It is worth noting that, although in this proof we used the universal cloner map \cite{keyl1999optimal}, 
the proof does not rely on the specific properties of this  map. For instance, rather than the universal cloner, we could have used a covariant measure-and-prepare map, which first performs a covariant measurement on input $A$, and then prepares states of $A_1 \cdots A_n$ according to the outcome of the measurement.  In this case, one can also show that $\sigma_{A_i}$ 
can be an asymmetric state, independent of the choice of $n$, and the above argument can be applied to prove the result.

\section{Proof that the function $f_t$  is a measure of translation asymmetry for any $t\in\mathbb{R}$}

For any $t\in \mathbb{R}$, we have defined
\beq\label{defnft}
f_t(\rho)\equiv1-\text{Fid}(\rho,e^{-i H t}\rho e^{i H t}),\eeq 
where $\text{Fid}(\rho,e^{-i H t}\rho e^{i H t})=\|\sqrt{\rho}\sqrt{e^{-i H t}\rho e^{i H t}} \|_1$ is the (Uhlmann) fidelity between states $\rho$ and  $e^{-i H t}\rho e^{i H t}$.

To see that $f_t(\rho)$ takes values in the range $[0,1]$, it suffices to note that the fidelity between any two quantum states is a value in the range $[0,1]$.

To prove that $f_t$ is a measure of translational asymmetry for any $t\in\mathbb{R}$, one must show (i) that it is zero for incoherent states, and (ii) that it is non-increasing under translationally covariant operations.  


To see that $f_t$ is zero for incoherent states, it suffices to note that for incoherent $\rho$, $[\rho, H]=0$, which immediately implies  $f_t(\rho)=0$ by Eq.~\eqref{defnft}.  

To see that $f_t$ is non-increasing under translationally covariant operations), it suffices to note that
for any translationanlly covariant operation $\mathcal{E}$,
\begin{align}
f_t(\mathcal{E}(\rho))&=1-\text{Fid}(\mathcal{E}(\rho),e^{-i H t}\mathcal{E}(\rho) e^{i H t})\\ &=1-\text{Fid}(\mathcal{E}(\rho), \mathcal{E}(e^{-i H t}\rho e^{i H t}))\\ &\le 1-\text{Fid}(\rho, e^{-i H t}\rho e^{i H t})\\
&= f_t(\rho)\ ,
\end{align}
where in the second line we have used the covariance of $\mathcal{E}$, \rob{in the third line we have used the monotonicity of the fidelity, and in the fourth line we have used the definition \eqref{defnft}.}


\section{Proof of the tradeoff relation (Equation 6 in the {letter})}


Recall that we have defined a degree of irreversibility of a state conversion $\rho_Q \to \sigma_{Q}$ by 
\beq\label{defnirrev}
\text{irrev}(\rho_Q,\sigma_{Q})\equiv 1-\max_{\mathcal{R}_Q}\ \text{Fid}^2\left(\rho_Q, \mathcal{R}_Q(\sigma_{Q})\right),
 \eeq
where $\mathcal{R}_Q$ is a covariant recovery operation.  Relative to this definition, and the definition of the measure of asymmetry $f_t$ in Eq.~\eqref{defnft}, the tradeoff relation we seek to prove here is as follows: if the state conversion 
$\psi_Q \rightarrow \sigma_{QS'}$ (where $\psi_Q$ is pure)
 is achievable by a translationally covariant operation, then
\beq\label{Eq-trade2}
\forall  t\in\mathbb{R}:\ \ \    f_t(\sigma_{S'})\le 4 \frac{\sqrt{\text{irrev}(\psi_Q,\sigma_{Q})}}{1-f_t(\psi_Q)}\ .
\eeq
If it can be shown that for {\em all} translationally covariant candidates for the recovery operation, $\mathcal{R}_Q$, it holds that
\beq\label{Eq-trade3}
\forall  t\in\mathbb{R}:\ \ \  \left(1-f_t(\psi_Q) \right)  f_t(\sigma_{S'})\le 4  \sqrt{ 1-\text{Fid}^2\left(\psi_Q, \mathcal{R}_Q(\sigma_{Q})\right) }\ .
\eeq
 then this equation also holds for the particular recovery operation that achieves the maximum value of $\text{Fid}^2\left(\psi_Q, \mathcal{R}_Q(\sigma_{Q})\right)$, and then the tradeoff relation follows directly from Eq.~\eqref{defnirrev} whenever $f_t(\psi_Q) <1$. 
It suffices, therefore, to establish Eq.~\eqref{Eq-trade3} for {\em all} translationally covariant operations $\mathcal{R}_Q$.


Consider the post-recovery state $\omega_{QS'}$ associated to the recovery operation $\mathcal{R}_Q$,
\beq
\omega_{QS'}\equiv (\mathcal{R}_Q\otimes \mathcal{I}_{S'}) (\sigma_{QS'})\ ,
\eeq
where $\mathcal{I}_{S'}$ is the identity operation on $S'$.  
We denote the marginals on $S'$ and $Q$ of the post-recovery state by $\omega_{S'} \equiv \Tr_Q(\omega_{QS'})$ and $\omega_Q \equiv \Tr_{S'}(\omega_{QS'})$ respectively. \color{black} Note that
\beq
\omega_{S'}= \Tr_Q(\omega_{QS'})=  \Tr_Q(\sigma_{QS'})=\sigma_{S'}\ .
\eeq
\color{black}
Using $\omega_Q$, we can rewrite Eq.~\eqref{Eq-trade3} as 
\beq\label{Eq-trade4}
\forall  t\in\mathbb{R}:\ \ \  \left(1-f_t(\psi_Q) \right)  f_t(\sigma_{S'})\le 4  \sqrt{ 1-\text{Fid}^2\left(\psi_Q, \omega_Q \right) }\ .
\eeq
This is what will be proven below.

The starting inequality for our proof is obtained by an application of the following lemma:
\begin{lemma}[\cite{marvian2018coherence}] \label{lem:fid}
For any pairs of states $\tau_1$ and $\tau_2$ and unitary $U$, it holds that
\bes\label{Bures}
\begin{align}
\Big|\textrm{Fid}(U \tau_1 U^\dag , \tau_1)- \textrm{Fid}(U \tau_2 U^\dag , \tau_2)\Big| &\le 4\sqrt{1-\textrm{Fid}(\tau_1,\tau_2)}\ .
\end{align}
\ees
\end{lemma}

Applying this lemma with $\tau_1$ as the post-recovery state $\omega_{QS'}$, $\tau_2$ as the initial state $|\psi\rangle\langle\psi|_Q\otimes \sigma_{S'}$ and $U$ as the time-translation $e^{-iH_{QS'} t}$, and recalling the definition of $f_t$ from Eq.~\eqref{defnft}, we obtain
\beq\label{fromlemma}
|f_t( \psi_Q\otimes \sigma_{S'})-f_t(\omega_{QS'})|\le 4\sqrt{1-\text{Fid}( \psi_Q \otimes \sigma_{S'}\ ,\ \omega_{QS'})}\ .
\eeq

It remains only to show that Eq.~\eqref{fromlemma} implies Eq.~\eqref{Eq-trade4}.

We begin with the left-hand side of Eq.~\eqref{fromlemma}. We note that by the definition of $f_t$ in Eq.~\eqref{defnft}, and the assumption that $Q$ and $S'$ are noninteracting ($H_{QS'}= H_Q + H_{S'}$), we have
\begin{align}\label{decompftcomp}
f_t(|\psi\rangle\langle\psi|_Q\otimes \sigma_{S'})&=1-\text{Fid}\Bigl(|\psi\rangle\langle\psi|_Q\otimes \sigma_{S'}\ ,\  U_Q(t)|\psi\rangle\langle\psi|_Q U^\dag_Q(t) \otimes U_{S'}(t) \sigma_{S'} U^\dag_{S'}(t) \Bigr)\\ &=1-\text{Fid}\Bigl(|\psi\rangle\langle\psi|_Q ,\  U_Q(t)|\psi\rangle\langle\psi|_Q U^\dag_Q(t)  \Bigr)\, \text{Fid}\Bigl(\sigma_{S'}\ , U_{S'}(t) \sigma_{S'} U^\dag_{S'}(t) \Bigr)\\ 
&=1-\left[1-f_t(\psi_Q)\right] \left[1-f_t(\sigma_{S'}) \right]\ .
\end{align}

It follows that the left-hand side of Eq.~\eqref{fromlemma} can be rewritten as
\beq\label{fromlemma2}
|f_t( \psi_Q\otimes \sigma_{S'})-f_t(\omega_{QS'})|
 = \Big|1-\left[1-f_t(\psi_Q)\right] \left[1-f_t(\sigma_{S'}) \right] - f_t(\omega_{QS'})\Big|
\eeq
Next, we note that
\beq\label{sdsds}
f_t(\psi_Q)\ge f_t(\omega_{QS'})\ 
\eeq
based on the monotonicity of $f_t$ under translationally covariant operations together with the fact that  $\omega_{QS'}$ is obtained from $\psi_Q$ via a translationally covariant operation.

This means that
\begin{align}
1-\left[1-f_t(\psi_Q)\right] \left[1-f_t(\sigma_{S'}) \right] - f_t(\omega_{QS'})&\ge [1-f_t(\psi_Q)] -\left[1-f_t(\psi_Q)\right] \left[1-f_t(\sigma_{S'}) \right]\\ &=[1-f_t(\psi_Q)] f_t(\sigma_{S'}) \\  &\ge 0\ ,
\end{align}
where the first inequality follows from Eq.~\eqref{sdsds}, and the second inequality follows from the fact that $f_t$ is in the range $[0,1]$.

Combing this with Eq.~\eqref{fromlemma2}, we obtain the following lower bound for the left-hand side of Eq.~\eqref{fromlemma},
\begin{align}\label{LHSbound}
|f_t( \psi_Q\otimes \sigma_{S'})-f_t(\omega_{QS'})|=\Big|1-\left[1-f_t(\psi_Q)\right] \left[1-f_t(\sigma_{S'}) \right] - f_t(\omega_{QS'})\Big| &\ge \Big| \left[1-f_t(\psi_Q)\right] f_t(\sigma_{S'}) \Big|.
\end{align}
Next, we turn to the right-hand side of Eq.~\eqref{fromlemma}.

It remains only to show that 
\beq
\text{Fid}(|\psi\rangle\langle\psi|_Q\otimes \sigma_{S'}\ ,\ \omega_{QS'})\ge \text{Fid}^2(|\psi\rangle\langle\psi|_Q \ ,\ \omega_{Q})= \langle\psi|\omega_{Q} |\psi\rangle\ ,
\eeq
\color{black}
or, equivalently, 
\beq\label{lastineq}
\text{Fid}(|\psi\rangle\langle\psi|_Q\otimes \omega_{S'}\ ,\ \omega_{QS'})\ge \text{Fid}^2(|\psi\rangle\langle\psi|_Q \ ,\ \omega_{Q})= \langle\psi|\omega_{Q} |\psi\rangle\ ,
\eeq
\color{black}
because this, together with Eq.~\eqref{LHSbound}, establishes what we need to show, namely, that Eq.~\eqref{fromlemma} implies Eq.~\eqref{Eq-trade4}.

Suppose  Eq.~\eqref{lastineq} holds in the special case where $\omega_{QS'}$  is a pure state. Then, using the fact that the fidelity is a concave function of each of its arguments, together with the fact that the right-hand side of Eq.~\eqref{lastineq} is linear in  $\omega_Q$, one can extend this result to the general case, where  $\omega_{QS'}$ is an arbitrary mixed state. 
Consequently, it suffices to prove Eq.~\eqref{lastineq} in the case of $\omega_{QS'}$ being pure.

Letting 
\beq
\omega_{QS'}=|\Omega \rangle\langle\Omega |_{QS'}.
\eeq  
and recalling that if one of the arguments of the fidelity is a pure state $\Psi = |\Psi\rangle\langle\Psi|$, then $\text{Fid}(\tau,\Psi) =\sqrt{\langle \Psi | \tau | \Psi\rangle}$, it follows that we must show  that 
\color{black}
\beq\label{wgggs}
\langle \Omega | \Big(|\psi\rangle\langle\psi|_Q\otimes \omega_{S'}\Big) | \Omega\rangle_{QS'} \ge 
\langle\psi|\, \omega_{Q} |\psi\rangle_Q^2.
\eeq

\color{black}

 Consider the decomposition of $|\Omega\rangle_{QS'}$ into its component {\em within} the subspace associated to the projector $|\psi\rangle\langle\psi|_Q \otimes \mathbb{I}_{S'}$ and its component orthogonal to this subspace.
 This can be written as
 \beq\label{Omega}
 |\Omega\rangle_{QS'}=\sqrt{p}  |\psi\rangle_Q \otimes |\eta\rangle_{S'} +\sqrt{1-p}  |\gamma\rangle_{QS'}\ ,
 \eeq
 where 
 \begin{align}\label{defnp}
 p&\equiv \langle\Omega|\Big( |\psi\rangle\langle\psi|_Q \otimes \mathbb{I}_{S'}\Big)  |\Omega\rangle_{QS'}\\
 &=\langle\psi|\omega_Q|\psi\rangle_Q\ ,
 \end{align}
and 
 \color{black}
 \beq
 |\eta\rangle_{S'} \equiv \frac{1}{\sqrt{p}}\Big( \langle\psi|_Q \otimes \mathbb{I}_{S'} \Big)  |\Omega\rangle_{QS'}\ ,
 \eeq
 and where $\sqrt{1-p} |\gamma\rangle_{QS'}$\color{black} is the component of $ |\Omega\rangle_{QS'}$ which is orthogonal to $ |\psi\rangle_Q\otimes |\eta\rangle_{S'}$.
 \color{black}
 
Eq.~\eqref{Omega} implies that
 \begin{align}\label{penul}
\langle\Omega|\Big( |\psi\rangle\langle\psi|_Q \otimes \omega_{S'}\Big) |\Omega\rangle_{QS'}
 &=p \langle\eta |\omega_{S'} |\eta\rangle_{S'}.
\end{align}
However, noting that
\begin{align}
\langle \Omega | \Big(\mathbb{I}_Q  \otimes  |\eta_{S'} \rangle\langle\eta|_{S'} \Big)  |\Omega\rangle_{QS'} \ge \langle \Omega |\Big(\color{black} |\psi\rangle\langle\psi|_Q \otimes |\eta_{S'} \rangle\langle\eta|_{S'}\color{black} \Big)  |\Omega\rangle_{QS'} \color{black} =p \color{black},
\end{align}
 and that 
 \beq
\langle\Omega|\Big( \mathbb{I}_Q \otimes |\eta\rangle\langle\eta|_S\Big)  |\Omega\rangle_{QS'} = \langle\eta|\omega_{S'}|\eta\rangle_{S'}. 
\eeq
it follows that 
\beq
 \langle\eta|\omega_{S'}|\eta\rangle_{S'} \ge  p.
\eeq
Substituting this into Eq.~\eqref{penul} and recalling the definition of $p$ from Eq.~\eqref{defnp}, we arrive at Eq.~\eqref{wgggs}.
This concludes the proof.

\section{Recovering no-broadcasting of asymmetry from the tradeoff relation for the case of pure states}

 
Here we prove that the special case of our no-broadcasting theorem wherein the state $\rho_Q$ is pure follows from our tradeoff relation, Eq.~\eqref{Eq-trade2}.

We take $\rho_Q = \psi_Q$ to denote the purity assumption.  Eq.~\eqref{Eq-trade2} implies that whenever $f_t(\psi_Q)<1$, if the state conversion  $\psi_Q\rightarrow \sigma_{Q}$ is reversible, so that $\text{irrev}(\psi_Q,\sigma_{Q})=0$, then $ f_t(\sigma_{S'})=0$.  However, because $f_t$ is not a faithful measure of asymmetry, this is not sufficient to infer that $\sigma_{S'}$ is symmetric.  However, this conclusion {\em does} follow if $ f_t(\sigma_{S'})=0$ for a finite neighbourhood around $t=0$, and the latter is the case whenever the expectation value of the generator of the symmetry, $H_Q$, on the initial state, $|\langle\psi_Q|H_Q|\psi_Q\rangle|$, is finite (i.e., $\psi_Q$ is of bounded-size).  This last inference follows from the fact that for the neighbourhood of $t=0$ defined by $\color{black}|t|\le |\langle\psi_Q|H_Q|\psi_Q\rangle|^{-1}$\color{black}, $\psi_Q$ cannot be perfectly distinguishable from $e^{-i H_Q t}\psi_Q e^{i H_Q t}$,  so that for this finite neighborhood, $f_t(\psi_Q)<1$, and therefore, by the tradeoff relation, $f_t(\sigma_{S'})=0$ as well.
  
%
%
%

\end{document}